\documentclass[
twocolumn,
showpacs,
a4paper]{revtex4} 

\usepackage{graphicx,xcolor}
\usepackage{amsmath, bbm}
\usepackage[applemac]{inputenc}
\usepackage{epstopdf}
\usepackage{bbm}

\newcommand{\mean}[1]{\langle #1\rangle}

\renewcommand{\Re}{\mbox{Re }}
\newcommand{\re}[1]{\mathrm{Re}\,#1}

\renewcommand{\imath}[0]{\mathrm{i}}

\newcommand{\unit}[1]{\,\mathrm{#1}}
\renewcommand{\vec}[1]{\boldsymbol{#1}}

\begin{document}
 \title{Reduction of nonclassical fluctuations by entangled nonidentical emitters in nanophotonic environments}
 \author{Harald R. Haakh}
 \email{harald.haakh@mpl.mpg.de }
 \author{Diego Mart\'in-Cano} 
  \email{diego-martin.cano@mpl.mpg.de }
  \affiliation{
Max Planck Institute for the Science of Light, G\"unther-Scharowski-Stra{\ss}e 1/24, D-91058 Erlangen, Germany. 
}

\date{\today}

\begin{abstract}
We propose a scheme in which broadband nanostructures allow to generate squeezed light and entanglement of quantum emitters that are extremely far detuned.
It is shown that the reduced fluctuations of the electromagnetic field arising from collective resonance fluorescence provide also a means to detect the entanglement between the emitters. 
Due to the near-field enhancement in the proposed hybrid systems, these nonclassical effects can be encountered outside both the extremely close separations limiting the observation in free space and narrow frequency bands in high-Q cavities.
Our approach permits to overcome the limitations of noninteracting single emitters and is more robust against phase decoherence induced by the environment.
\end{abstract}


\maketitle

The generation of entanglement, one of the most interesting and nonintuitive resources in quantum optics, lies at the heart of every physical process that surpasses classical limits.
Entanglement between quantum emitters is necessary to perform nonclassical information processing \cite{Bremner2002} and to go beyond the standard sensitivity limits in interferometry \cite{pezze2009}.
Nonclassicality can also be encountered in squeezed states of light \cite{Loudon1987} and becomes manifest in the reduced quantum fluctuations of the electromagnetic field below the shot-noise level that provide the key to the realization of quantum-enhanced applications in communication, spectroscopy \cite{Drummond2010}, and imaging \cite{Taylor2014}.
The close connection between different manifestations of quantum correlations such as entanglement and squeezed states in spin-ensembles \cite{pezze2009}, also opens interesting questions regarding the transfer of entanglement in matter to squeezing in light \cite{Saito1997, Grunwald2014}
in order to share its nonclassical properties for quantum applications \cite{Meyer2001}.
Despite extensive studies of the creation of entanglement between emitters by means of continuous-wave squeezed light \cite{Hammerer2010,Ficek2002}, 
the possibility to characterize entanglement via the detection of squeezed light remains widely unexplored. 

So far, the emission of squeezed light in resonance fluorescence \cite{Lu1998,Ourjoumtsev2011} and the generation of entanglement between quantum emitters \cite{Hagley1997} have been demonstrated in a fundamental manner, mainly in free space or microcavities \cite{Ficek2002}.
Recently, there is a strong interest in exploring these quantum effects in new regimes in the coupling strength between matter and light by bringing emitters to integrated nanophotonic systems \cite{Lodahl2013,Tame2013}. 
The strength of nanostructures relies on their ability to modify both the electromagnetic near and far-field \cite{Novotny2006}, which allows for enhancing and controlling the interaction between emitters \cite{Dung2002}. 
Several schemes for such hybrid environments have been suggested for the generation of entanglement between quantum emitters 
\cite{Lodahl2013,Gonzalez-Tudela2011,Gullans2012,Gonzalez-Tudela2013,Tame2013,Lee2013,Hou2014}, whereas only recently a nanoscopic source of squeezed light has been proposed 
\cite{Martin-Cano2014}.
At present, the experimental observation of entanglement in nanophotonic systems remains elusive.  One remaining difficulty is that most of the proposed mechanisms rely on either interference arising in detection with low generation probabilities \cite{Lee2013}, or on the availability of almost identical emitters \cite{Gonzalez-Tudela2011,Gullans2012,Gonzalez-Tudela2013}.
The latter is notoriously difficult in common solid state emitters, as local properties of the host matrix inhomogeneously distribute the optical transitions, typically by several $\unit{GHz}$ \cite{Hettich2002, Faez2014,Patel2010}.
These lie outside the reach of narrowband cavities and thus require sophisticated tuning methods.

In this Letter we show how broadband nanostructures assist in the generation of squeezed light arising from the deterministic entanglement of two  far detuned emitters. 
We identify the connection of the entanglement between the emitters and the reduced light fluctuations in cooperative resonance fluorescence assisted by nanoarchitectures.
%
\begin{figure}[t!!]
\raggedright
 \includegraphics[width=.95\columnwidth]{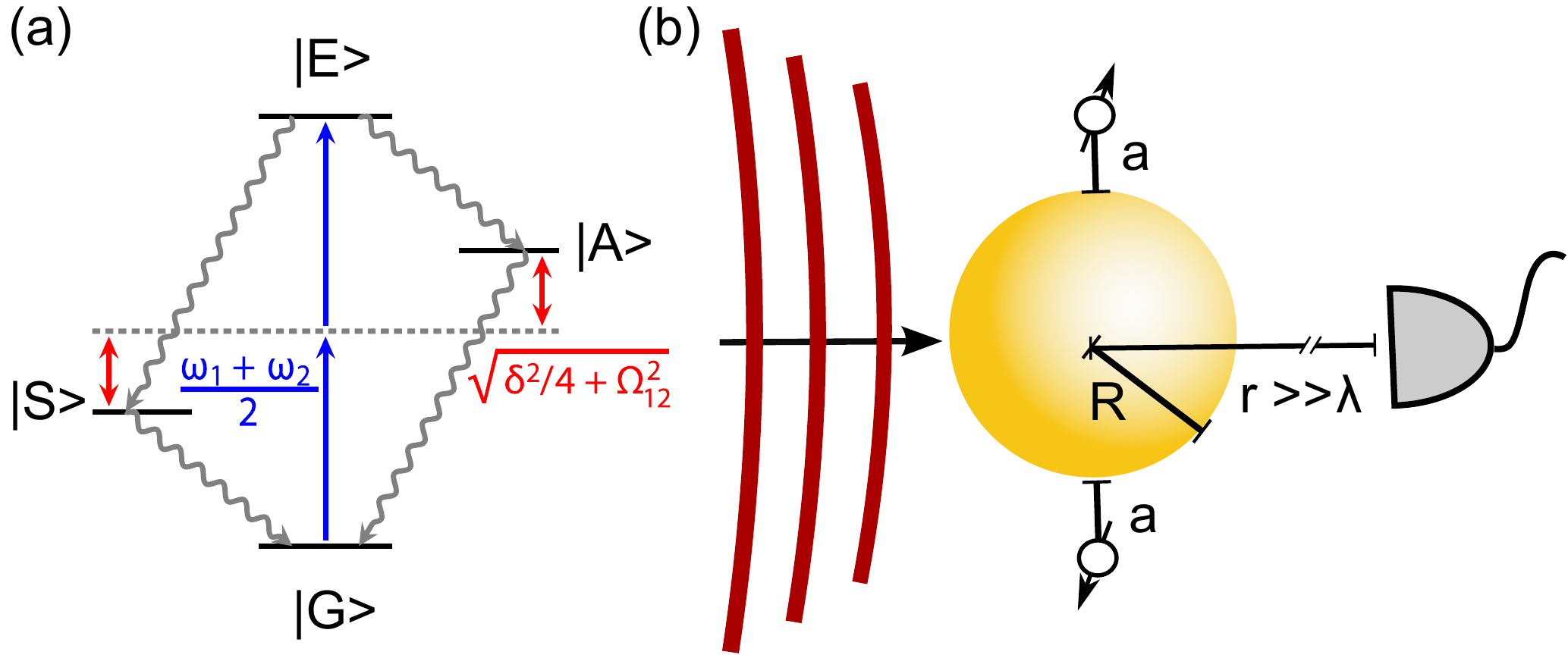}\\
\includegraphics[width=\columnwidth]{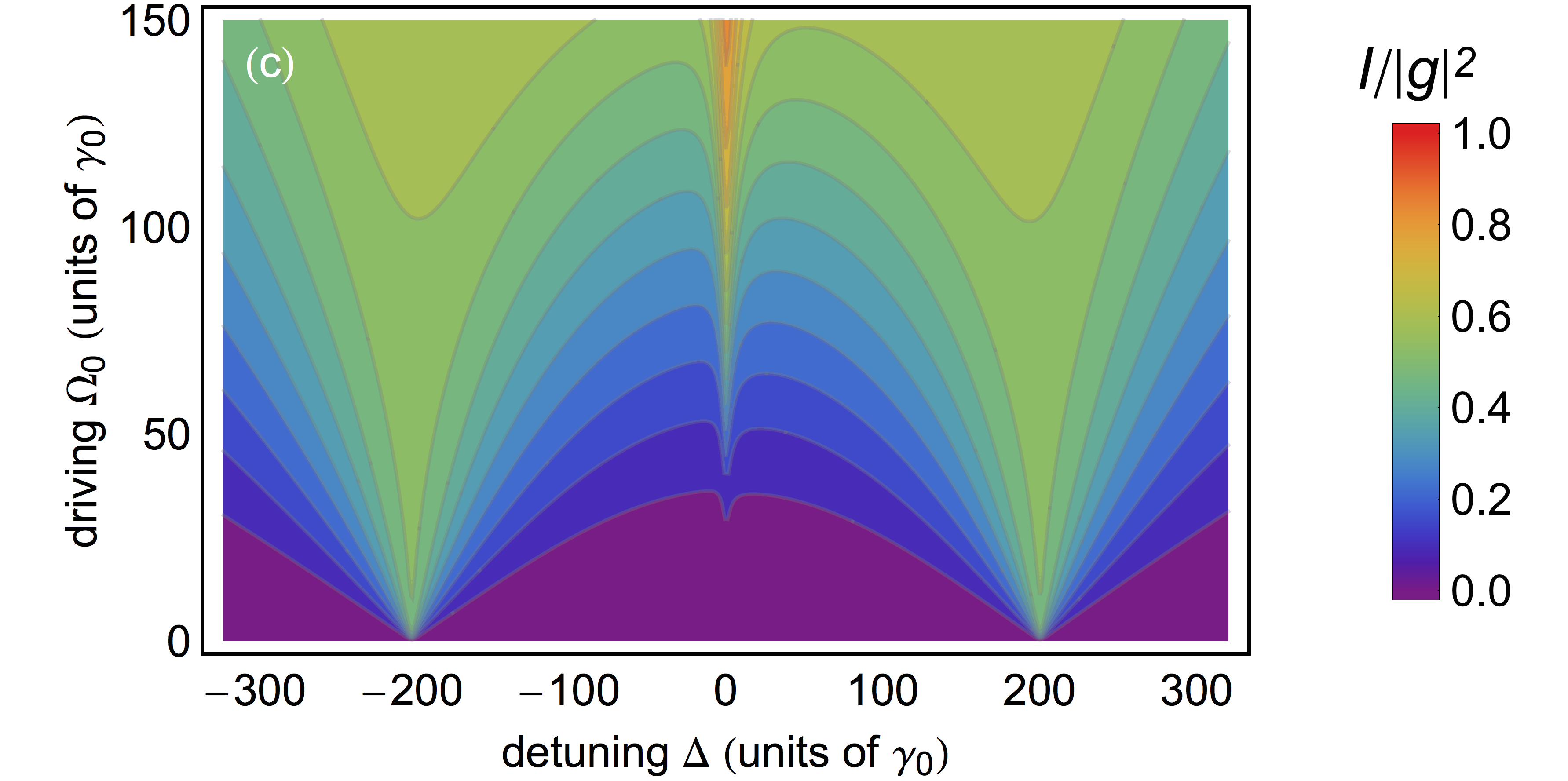}
\caption{ a) Level scheme of the coupled system, consisting of the fundamental state $|G\rangle$ and  doubly excited state $|E\rangle$, connected by a nonlinear two-photon process $\Delta = \omega_L - (\omega_1 + \omega_2)/2 =0$ (blue arrow).
Due to the emitter detuning $\delta$ and coupling $\Omega_{12}$, the dark and bright single-excitation states $|A\rangle$ and $|S\rangle$ are nondegenerate.
b) Schematic setup consisting of two emitters coupled to a nanoantenna and illuminated by a coherent driving field. The observation takes place in the far field.
c) Incoherent fluorescence signal $I=\sum_i\langle \vec{E}^{-}_i\cdot\vec{E}^{+}_i\rangle$ as a function of the driving field detuning $\Delta$ and free-space Rabi frequency $\Omega_0$.
The configuration of two detuned emitters ($\lambda = 780\unit{nm}, \delta = 400\gamma_0, \Omega_{12} = -6.4 \gamma_0, \gamma_{12} = -2.6 \gamma_{0}, \gamma = 2.9 \gamma_0$) corresponds to distances $a=25~\unit{nm}$ from a $R=40\,\unit{nm}$ gold nanosphere.
The local driving field enhancement near the nanosphere by a factor $\Omega\approx 2\Omega_0$ is taken into account.
}
\label{fig:1}
\end{figure}
%
Our scheme is based on a two-photon process in a coupled pair of two-level emitters with transition frequencies separated by many linewidths ($\omega_2-\omega_1= \delta\gg \gamma$).
Fig.\,\ref{fig:1}a illustrates the four-level structure of the coupled system.
%
Two-photon transitions from the fundamental state $|G\rangle$ to the doubly excited state $|E\rangle$  at the midfrequency $(\omega_1 + \omega_2)/2$ (blue arrows) become allowed as a result of the dipole-dipole coupling potential $\hbar \Omega_{12}$. For  emitters coupled by the extreme near field in a bulk medium \cite{Varada1992}, incoherent nonlinear transitions have been verified experimentally \cite{Hettich2002}.
Here, we exploit the coherent regime of the transition. This is possible when $\Omega_{12}$ becomes comparable to the emitters' decay rate $\gamma$ and when the driving fields, characterized by the Rabi frequency $\Omega_{}$ and the laser detuning $\Delta = \omega_L - (\omega_1+\omega_2)/{2}$, are weak.
The coherence, however, degrades in stronger fields, when the doubly excited state is populated via collective transitions ($\propto\Omega^4 \Omega_{12}$) or -- to a lesser extent -- single-photon transitions ($\propto\Omega^4 \Delta$) \cite{Varada1992}.
Enhanced dipole-dipole coupling \cite{Agarwal1998,Gonzalez-Tudela2011} is in reach outside the extreme near-field regime of bulk environments, in state-of-the-art subwavelength broadband cavities \cite{Kelkar2015}, nanowaveguides \cite{Lodahl2013,Faez2014}, or antenna structures \cite{Tame2013}.
All these hybrid systems can provide the necessary interaction to overcome saturation broadening and permit much larger emitter separations or even far-field coupling.
We find that the nanostructure-assisted nonlinearity allows for the generation of entanglement and, in consequence, the emission of squeezed light, exceeding the performance of uncoupled emitters.

As a proof of principle, we consider the system depicted in Fig.\,\ref{fig:1}b, consisting of two quantum emitters placed at a distance $a=25\unit{nm}$ from a subwavelength gold nanosphere of radius $R = 40\unit{nm}$ and symmetrically driven. 
%
Each of the emitters ($i = 1,2$) is characterized by the coherence operator of a two-level emitter, $\hat \sigma_i = |g\rangle\langle e|$ with resonance frequencies separated by $\delta = 400 \unit{\gamma_0}$.
Near the nanostructure, not only the dipole-dipole coupling $\Omega_{12}$ and the incoherent coupling rate $\gamma_{12}$, but also the emitter linewidth $\gamma$ and Rabi frequency $\Omega$ are altered from their values in free space, $\gamma_0$ and $\Omega_0$, increasing the bandwidth of the interaction and requiring lower driving powers.
The positive frequency fields $\hat{\vec{E}}^{+}_i = \vec{g}_i\hat{\sigma}_i$, scattered by each of the emitters in the presence of the nanostructure to the far-field, can be expressed in terms of the classical Green's tensor. 
We then obtain the steady-state correlations of the total scattered field from the quantum optical master equation (see Supplemental Material for details).

Since the coherent dynamics are limited by the population of the excited states, we first characterize the response to the driving field in terms of the incoherent fluorescence signal $I= \sum_i \langle\hat{\vec{E}}^{-}_i \cdot \hat{\vec{E}}^{+}_i\rangle$, which is a common experimental measure of the excitation spectrum  and allows to identify the dipole-dipole coupling \cite{Hettich2002}.
For relatively equivalent positions of the emitters and a balanced detector position, we may assume both equal scattering amplitudes and decay rates $\gamma_i = \gamma$ (the impact of asymmetric configurations is discussed in the Supplemental Material).
Fig.\,\ref{fig:1}c displays the incoherent fluorescence as a function of the laser detuning $\Delta$ and the free-space driving amplitude $\Omega_0$.
Its onset suggests a rough boundary of the coherent regime. 
Apart from the single-photon resonances at $\Delta \approx \pm\sqrt{(\delta/2)^2 + \Omega_{12}^2}$, there is a clear nonlinear absorption peak at $\Delta=0$ that arises from the direct two-photon transition from the fundamental state $|G\rangle$ to the doubly excited state $|E\rangle$.
The peak saturates for stronger driving $\Omega$, where it eventually overlaps with the saturation-broadened single-emitter lines.  
%
\begin{figure}
 \includegraphics[width=\columnwidth]{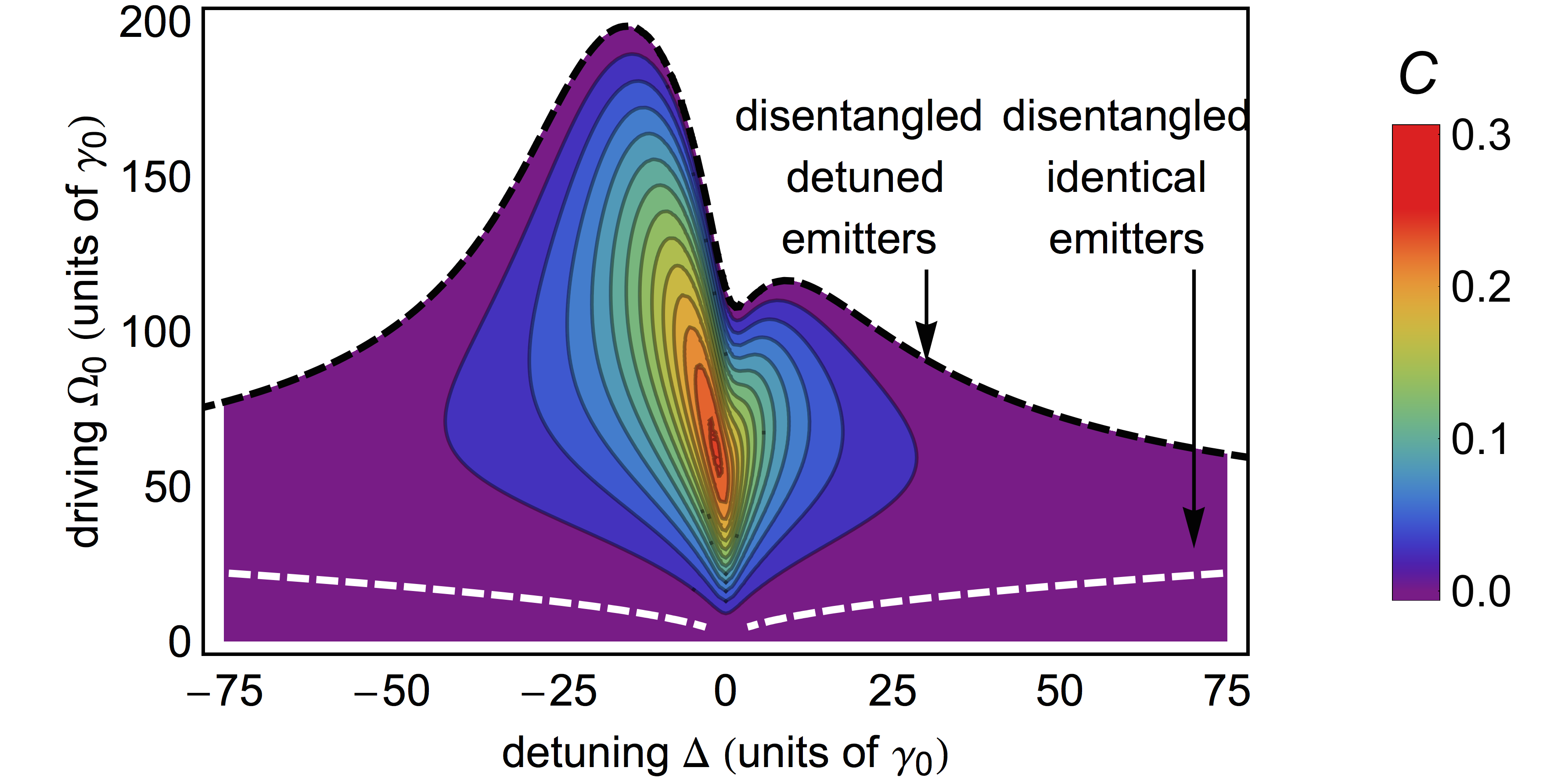}
\caption{Steady state concurrence as a function of the driving field amplitude and detuning. Parameters as in Fig.\,\ref{fig:1}.  Bipartite entanglement is excluded in the white area for detuned emitters ($\Delta > \Omega_{12}$) and is restricted to the region below the white dashed curve for identical emitters ($\Delta \ll \Omega_{12}$).}
\label{fig:2}
\end{figure}
%
The collective fluorescence peak at $\Delta=0$ would not be observable at comparable emitter separations and detuning in free space since the required stronger excitation field would mainly increase the single-emitter background.
   Moreover, the intensity  observed in the far field ($\propto|g|^2$) is enhanced by a factor $\approx1.7$ with respect to free space \cite{Martin-Cano2014} and it can still be significantly improved in engineered structures \cite{Tame2013,Rogobete2007a}.
 We finally notice that the large emitter detuning considered here would usually lie outside the interaction range of solid-state emitters in ultrahigh-Q cavities. For example, for cryogenic molecules with typical optical linewidths $\gamma_0/2 \pi\sim 50 \unit{MHz}$ \cite{Hettich2002}, the equivalent detuning $\delta/2\pi=20 \unit{GHz}$ would barely lie within the bandwidth of a cavity of $Q=10^4$.

The coherent excitation via the nanostructure-assisted dipole-dipole interaction makes it possible to generate steady-state entanglement between the two emitters despite their large detuning. We identify bipartite entanglement in the following by nonzero values of the concurrence  $C$, which ranges  between $0$ for fully separable states and $1$ for maximally entangled states \cite{Wootters1998}.
Fig.\,\ref{fig:2} shows an evaluation of the steady-state concurrence as a function of the driving field detuning and amplitude for the previous configuration.
 A significant amount of entanglement can be reached close to the two-photon resonance. As $\Omega$ is increased, the onset of saturation in the collective transition shifts this maximum in frequency (asymmetrically due to unequal populations of the entangled states $|A\rangle$ and $|S\rangle$), and finally suppresses entanglement.
 This behavior differs strongly from the case of identical emitters ($\delta \ll \Omega, \gamma$), for which single-emitter saturation suppresses the generation of entanglement at comparable driving $\Omega$ (in the region above the dashed white curve in Fig.\,\ref{fig:2}).
Moreover, the maximum concurrence $C \approx 0.3$ is comparable to  typical values  for identical emitters coupled via nanostructures \cite{Gonzalez-Tudela2011}.

%
Quantum correlations between the emitters \cite{Ficek2002, Tanas2003c, Saito1997} strongly affect the statistical properties of the scattered light. These can enhance squeezed resonance fluorescence \cite{Ficek1994} that is characterized by quadrature fluctuations below the shot-noise limit and can be experimentally addressed in homodyne or heterodyne setups \cite{Martin-Cano2014,Loudon1987}. 
For a general electric field quadrature 
\mbox{$\hat{\vec{E}}_{\theta}=\sum_{i}e^{i \theta}\hat{\vec{E}_i}^{+} + h.c.$}, 
 we identify reduced fluctuations of a single vector component by a negative value of the normally ordered variance
$
(\Delta\mathcal{E})^2 =\mean{:\!\!(\hat{E}_{\theta} - \mean{ \hat{E}_\theta})^2\!\!:}\break
=(\Delta\mathcal{E}_1)^2
	+(\Delta\mathcal{E}_2)^2
	+(\Delta\mathcal{E}_{12})^2
$. 
Here, the fluctuations from single emitters correspond to
\begin{align}	
\label{eqn:squeeze2}
\frac{(\Delta\mathcal{E}_i)^2}{2|g_i|^2}
&= 
	\biggl(
		\mean{\mathbf{\hat{\sigma}}_i^{\dag}\mathbf{\hat{\sigma}}_i}
		-|\mean{\mathbf{\hat{\sigma}}_i}|^2\biggr)
		\!-\!\textrm{Re}\left[
			e^{2 \imath (\theta +\phi_i)}\mean{\mathbf{\hat{\sigma}}_i}^{2}
		\right]	
\end{align}	
 with scattering phases $\phi_i = \arg(g_i)$ and recover previous results \cite{Martin-Cano2014} in the absence of interemitter coupling. The interaction modifies each of these terms and gives rise to the additional cross-correlations
\begin{align}
\label{eqn:squeeze3}
\frac{(\Delta\mathcal{E}_{12})^2}{ 2 |g_1 g_2| }	= 2\textrm{Re}\biggl[
	&e^{\imath  (\phi_1-\phi_2)}
		\biggl(
					\mean{\mathbf{\hat{\sigma}}^{\dag}_2\mathbf{\hat{\sigma}}_1}
					-\mean{\mathbf{\hat{\sigma}}^{\dag}_2}\mean{\mathbf{\hat{\sigma}}_1}
		\biggr)
	\\+&
	e^{\imath  (2 \theta+\phi_1+\phi_2)}
		\biggl(
					\mean{\mathbf{\hat{\sigma}}_2\mathbf{\hat{\sigma}}_1}
					-\mean{\mathbf{\hat{\sigma}}_2}\mean{\mathbf{\hat{\sigma}}_1}
		\biggr)
	\biggr]~
\nonumber
\end{align}
that clearly vanish for uncorrelated emitters.
Therefore, negative values arising from the cross-terms denote reduced fluctuations that allow to  identify bipartite entanglement.

\begin{figure}
\raggedright
 \includegraphics[width=\columnwidth]{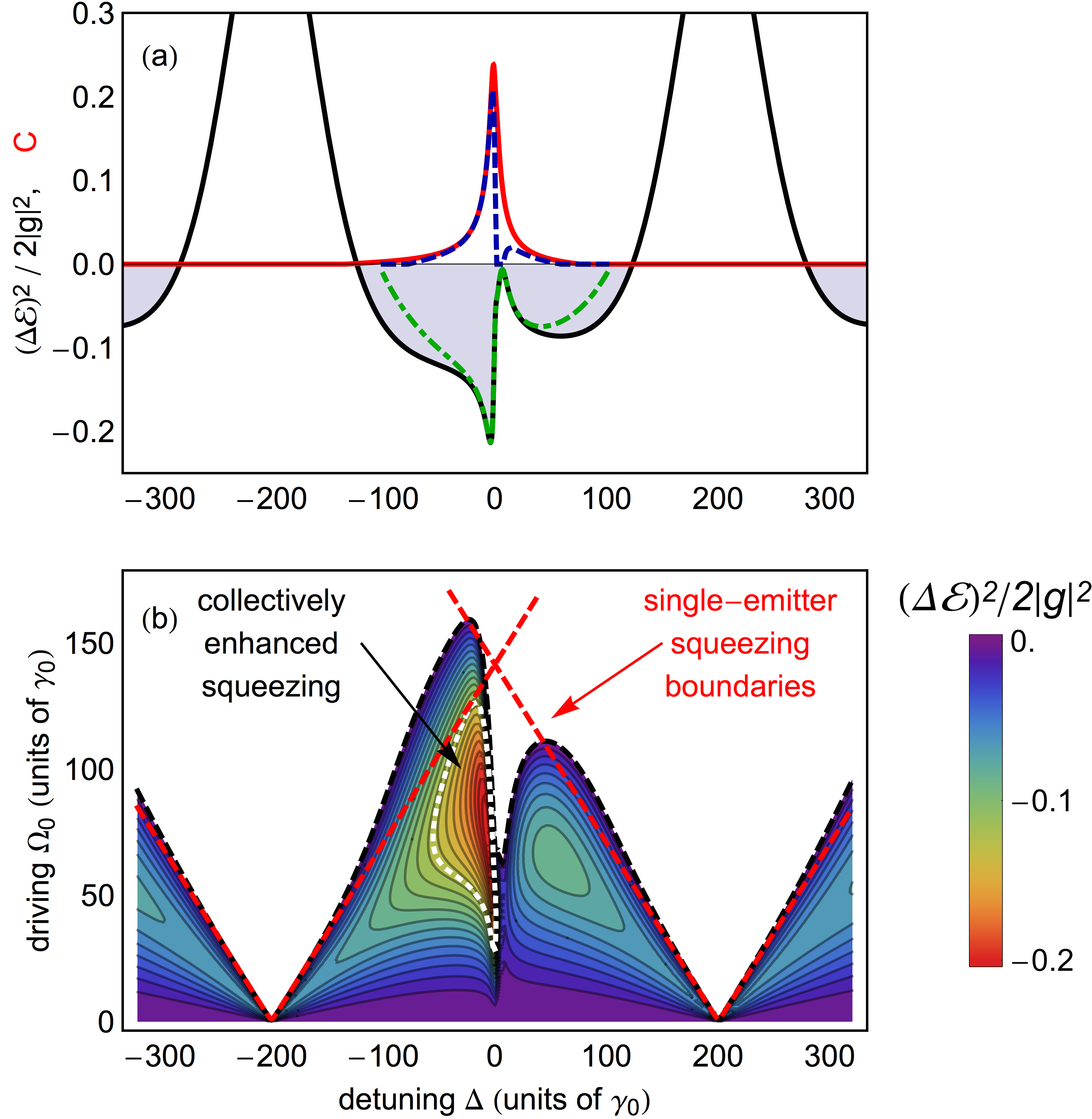}
  \caption{
a) Electric field fluctuations (black curve) obtained by optimizing the detection quadrature angle and steady state concurrence (red curve) vs. laser detuning at $\Omega_0 = 60 \gamma_0$.
The gray shading indicates squeezed light. The green and blue dashed curve gives the effective two-level approximation on the $|E\rangle-|G\rangle$-transition for the squeezing, and the concurrence, respectively.
b) Electric field fluctuations as a function of the driving field detuning and amplitude, optimizing the detection quadrature angle. No squeezing can be observed above the black dashed limit. The red curves indicate the limits for independent emitters. The fluctuations exceeds the minimum value feasible from two independent emitters inside the white contour. 
Parameters are chosen as in Fig.\,\ref{fig:1}.
}
\label{fig:3}
\end{figure}

The black curve in Fig.\,\ref{fig:3}a shows the optimum degree of squeezing $(\Delta \mathcal{E})^2/2|g|^2$, obtained by varying the phases $\theta, \phi_i$, as a function of the laser detuning.
The driving $\Omega_0 = 60\gamma_0$ was chosen well below the two-photon saturation.
As before, we focus on a favorable balanced detection configuration ($|g_{i}| = |g|$, see Supplemental Material for a discussion of the general case).
A distinct minimum of the fluctuations arises close to the two-photon resonance $\Delta=0$ and mirrors the behavior of the concurrence, indicated by the red (upper) curve. 
To exemplify the link between entanglement in matter and squeezing in light and to understand the main mechanism underlying the present scheme, we consider a simplified model for weak driving on the two-photon resonance.
An effective two-level picture arises, because the single-photon transitions are far detuned and the dipole-dipole coupling mainly induces a coherence $\rho_{EG}$ between the fundamental and doubly excited state, so that the density matrix is dominated by this element and the populations. From these elements, we find from Eqs.\,\eqref{eqn:squeeze2} and \eqref{eqn:squeeze3}  the normalized degree of squeezing
\begin{align}
\label{eq3}
\frac{(\Delta \mathcal{E})^2}{2|g|^2} \!\approx \! 1-\rho_{GG} \!+\! \rho_{EE} \!+\! \re\!\!\left[e^{\imath [\theta+\phi_1+\phi_2]}\rho_{EG}\right],\!
\end{align}
which is optimized when the last term equals $ - |\rho_{EG}|$.
The approximation (green dash-dotted curve) agrees excellently with the full squeezing amplitude (black curve).
Interestingly, Eq.\,\eqref{eq3} is closely related to the spin squeezing \cite{pezze2009, Hammerer2010} present in the system and reinforces the link to entanglement.
In fact, a similar treatment of the concurrence near the resonance  (blue dashed line) is equally dominated by $|\rho_{EG}|$ (see Supplemental Material for details).
Altogether this identifies the two-photon coherence as the source of both squeezing and entanglement and supports the description in terms of an effective two-level model.

A full evaluation of the electric field fluctuations as a function of the driving field amplitude and detuning is given in Fig.\,\ref{fig:3}b. At the chosen parameters, the strongest squeezing is encountered at $\Omega_0 \approx 75 \gamma_0$ near the center frequency $\Delta\approx0$ and closely follows the regime that allows for maximum bipartite entanglement in Fig.\,\ref{fig:2}. When the driving field intensity is further increased, saturation of the two-photon transition suppresses squeezing at $\Delta=0$, similar to the concurrence.
As the collective interaction also affects Eq.\,\eqref{eqn:squeeze2}, the boundaries of squeezed light can surpass the ones for uncoupled emitters \cite{Martin-Cano2014} (denoted by the red dashed limits in Fig.\,\ref{fig:3}a).
The strongest squeezing with values $(\Delta \mathcal{E})^2 /(2|g|^2)\approx -0.21$ significantly exceeds the universal threshold $-0.125$ that limits the squeezed resonance fluorescence from independent emitters \cite{Martin-Cano2014} and is indicated by the white contour in Fig.\,\ref{fig:3}b.
Thus, a strong collective enhancement of the squeezing per emitter \cite{Ficek1994} can be achieved despite their large detuning.
This is a general result for two-level systems coupled by means of strong dipole-dipole interactions and can be used to identify entanglement. 

\begin{figure}[t!]
\includegraphics[width=\columnwidth]{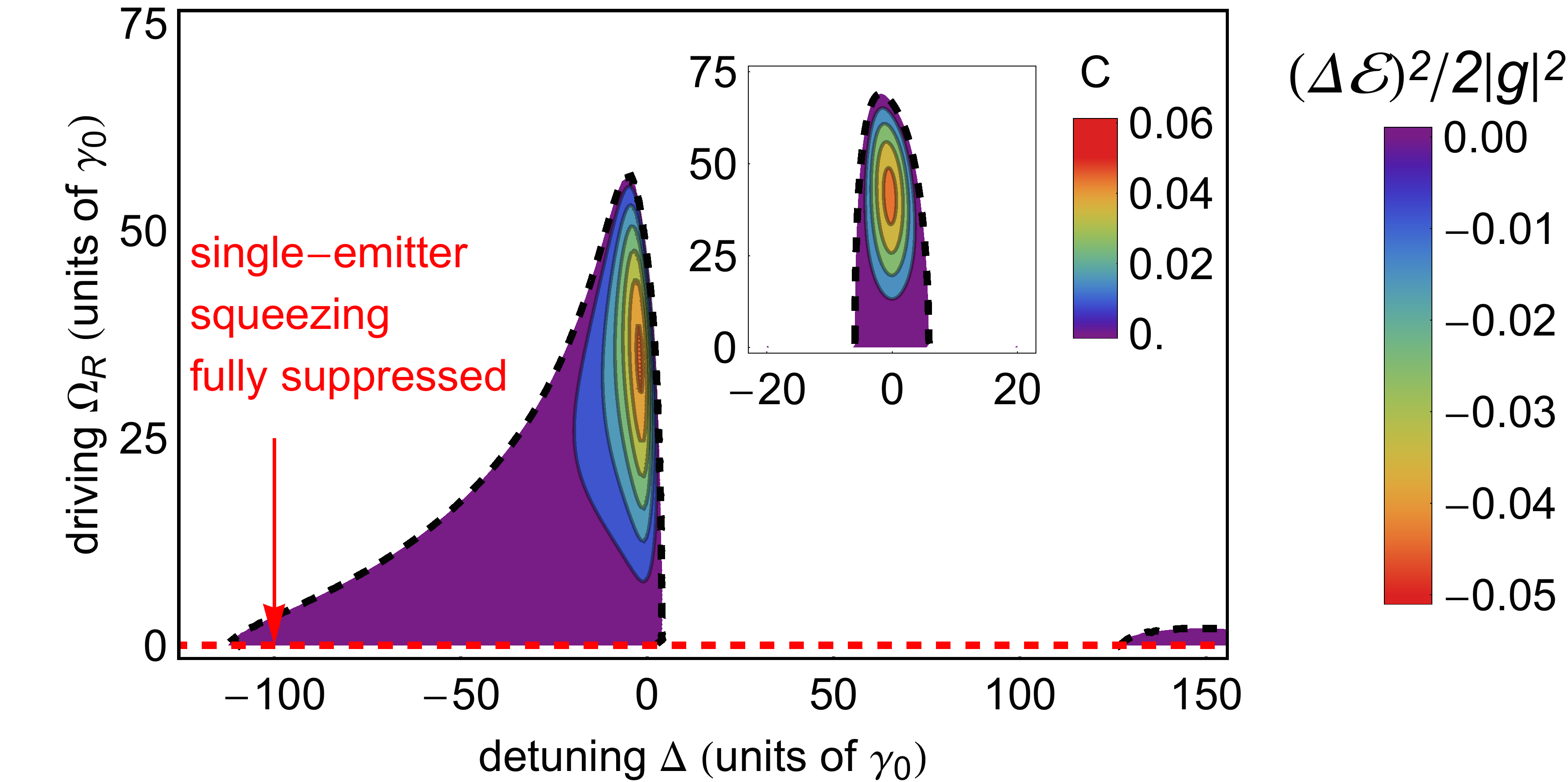}
 \caption{
Maximum degree of squeezing as a function of the driving field detuning and amplitude, obtained by optimizing the detection quadrature angle in the presence of strong additional pure dephasing processes at a rate $\gamma^* = 2 \gamma$ for which single-emitter squeezing is fully suppressed (cf. red dashed limit). 
 All other parameters as panel Fig.\,\ref{fig:1}. Inset: Analogous representation for the concurrence.
}
\label{fig:4}
\end{figure}
As an advantage for the realization of nanoscopic sources of squeezed light,
the collective effects induced by the nanoarchitecture allow for overcoming significant additional pure dephasing \cite{Carmichael2002}.
This phenomenon occurs commonly in solid-state environments due to the coupling to thermal phonons and it presents a main limitation to the operation time of quantum gates.
For independent emitters, additional pure dephasing at a rate $\gamma^* \ge \gamma/2$ will fully suppress the generation of squeezed light, even though the enhancing effect of a nanostructure may modify the boundary significantly \cite{Martin-Cano2014}.
At the onset of this unfavorable scenario, we represent in Fig.\,\ref{fig:4} the minimized electric field fluctuations  as a function of the driving field and detuning.
The full suppression of squeezing for uncoupled emitters is indicated by the red dashed limit, i.e. $(\Delta \mathcal{E})^2 \ge 0$ for all $\Omega_0$.
Still, a significant amount of squeezing survives near the two-photon resonance, though at a reduced absolute value. 
The optimum follows even more closely the behavior of the concurrence (shown in the inset) than in the absence of pure dephasing because terms stemming from the single emitters are now suppressed. These facts reflect that squeezed emission identifies entanglement despite additional dephasing and underline that nanostructures allow to taylor collective interactions, providing more robust sources of squeezed light in the presence of decoherence.

The combination of nanophotonic structures with quantum emitters presents a promising prospect for novel integrated sources of squeezed light and the generation of multipartite entanglement at the nanoscale.
The strong local field enhancement in nanoarchitectures feasible over a broad frequency range can provide significant intermitter coupling outside the regimes of extreme near fields in bulk media.
Using a metallic nanosphere as a paradigmatic example, we have found that both entanglement and squeezing can be achieved for two emitters that are mutually far-detuned, beyond the limits that are spectrally in reach in typical narrow-band cavities.
This can greatly relax the necessity of tuning precisely the optical resonances of solid-state emitters required to achieve large degrees of entanglement in hybrid systems and it can assist the future detection of entanglement.
Moreover, we have shown that the generation of squeezed light from interacting quantum emitters can serve as a reliable witness of entanglement.
Therefore, resonance fluorescence from interacting emitters can help to explore the transfer of nonclassicality from matter to light
and the distribution of useful quantum states in photonic devices that surpass the limits of classical optics \cite{pezze2009,Meyer2001}.
Our results show that collective interactions mediated by nanostructures can allow to overcome a significant amount of additional pure dephasing, even beyond the improvement achievable in independent-emitter squeezing in a nanostructure \cite{Martin-Cano2014}.
As both effects counteract phase decoherence -- a fundamental limit in setups at nonzero temperatures -- nanostructures can therefore allow to scale up the operating temperatures of solid-state emitters. A simple estimation for typical cryogenic molecules \cite{Irngartinger1997} shows that an increase of 300 in the radiative linewidth could allow to work at temperatures $\approx 70\unit{K}$ rather than below $5\unit{K}$, which would have significant technological implications.
Due to the generality of the formalism, our results may be transferred to other broadband environments such as optical antennas, subwavelength cavities, or nanowaveguides, where high coupling strengths may allow to push the limits for the generation of squeezed light even further.
In addition, we expect that the collective effects at work can be scaled up to higher numbers of emitters \cite{Gullans2012,Gonzalez-Tudela2013,Delga2014}, a future study of which will also provide a better understanding of the origin of squeezing in structured nonlinear media.
\acknowledgments
We thank A. Maser, B. Gmeiner, S. G\"otzinger, A. Gonz\'alez-Tudela, M. Agio, and V. Sandoghdar for helpful discussions.

 \section*{Supplemental Material}

\subsection{Master equation in macroscopic electrodynamics}
We use the approach of macroscopic quantum electrodynamics \cite{Knoll2001} to describe the behavior of $N$ two-level quantum emitters in an arbitrary environment. Each emitter is described by its transition dipole element $\vec{d}_n$, emitter operator $\hat{\sigma}_n$, and transition frequency $\omega_n$. 
Within the dipole, rotating wave, and  Markov approximation the full positive frequency field operator is given by
\begin{align}
\label{eq:field}
\boldsymbol{\hat{E}}^{+}_\mathrm{tot}(\mathbf{r},t)
	&= \boldsymbol{\hat{E}}^{+}_{\textrm{free}}(\mathbf{r},t)+\sum_{n=1}^N\boldsymbol{\hat{E}}^{+}_n(\mathbf{r},t)\\
%
\text {with }
%
\hat{\boldsymbol{E}}^{+}_{n}(\mathbf{r},t) 
	&=  \boldsymbol{g}_{n}(\vec{r})  \hat{\sigma}_n(t)~.
\label{eq:scattered_field}
\end{align}
Here, $\boldsymbol{g}_{n}(\vec{r}) \approx \frac{\omega_0^2}{c^2 \varepsilon_0} \boldsymbol{G}(\vec{r},\vec{r}_n, \omega_0) \cdot \vec{d}_n$  encodes the photon propagation in an arbitrary broadband environment via the  classical dyadic Green's tensor, where $\vec{r}$ denotes the detector position.
As long as dispersion in a nanostructure is weak, we can evaluate the Green's tensor at the center frequency $\omega_0 = \sum_{n} \omega_n / N$.
Since we consider far-field observation, we neglect a small position-dependent quantum correction connected with dispersion potentials, which would require some care in the study of near-fields \cite{Martin-Cano2014}. 
For a gold nanosphere, we evaluate the Green's function semi-analytically ~\cite{Dung2001,Li1994}, using a Drude-Lorentz interpolation of tabulated optical data for gold \cite{Lide2006}.

Under these conditions, the evolution equation is expressed in the interaction picture with respect to the free quantum fields and in the rotating frame of a classical driving field $E_{L}(\vec{r})e^{-\imath \omega_L t}+h.c.$ that translates to single-emitter Rabi frequencies $\Omega_{n} =  E_{L}(\vec{r}_n) \cdot \vec{d}_n/\hbar$. The Hamiltonian reads
\begin{align}
H = \hbar \sum_{n} \left[(\omega_n - \omega_L) \sigma_{z,n} -  \left( \frac{\Omega_{n}}{2} \sigma^+_n + h.c.\right)\right]~
\label{eq:hamiltonian}
\end{align}
with $\sigma_{z,n}$ the population inversion of particle $m$.
The quantum optical master equation in the Lindblad form is given by (see, e.g., Refs.\,\cite{Ficek2002,Dung2002a})
\begin{align}
\label{eq:master-equation}
\dot{\rho} &= - \frac{\imath}{\hbar} [(H -    \sum_{m\ne n} \Omega_{mn} \sigma_m^\dagger \sigma_n ), \rho] + \mathcal{L}[\rho]~,\\ 
\mathcal{L}[\rho] &= 
\sum_{m,n} -\frac{\gamma_{mn}}{2}\left(\sigma_m^\dagger \sigma_n \rho + \rho \sigma_m^\dagger \sigma_n - 2 \sigma_m \rho \sigma_n^\dagger \right)~.
\end{align}
Photon-mediated interactions result in modified dipole-dipole potentials $\Omega_{mn}$ while the dissipative dynamics, i.e. the individual and cooperative decay of dipole excitations, are described by the Liouvillian $\mathcal{L}$.
Here, the rates are the real and imaginary parts of  %
\begin{align}
\label{eq:decay_rate_frequency_shift}
\Omega_{mn} + \imath {\gamma_{mn}}/{2} ~=~& \vec{d}_m \cdot \vec{g}_n(\mathbf{r}_m).
\end{align}
Due to the local field enhancement, the single-body decay rates in the nanostructure  $\gamma_m\equiv\gamma_{mm}$ generally differ from the free-space value $\gamma_0$. For this work, the master equation \eqref{eq:master-equation} is solved numerically for the steady state solution in the case of $N=2$.
Aditional pure dephasing at a rate $\gamma_m^*$ can be included phenomenologically \cite{Carmichael2002} by adding the term 
\begin{align} 
\label{eq:puredephasing}
\mathcal{L}_{\rm d}[\rho] = 
\sum_{m} &-\gamma_m^*(\sigma_{z,m} \sigma_{z,m} \rho + \rho \sigma_{z,m} \sigma_{z,m}
 - 2 \sigma_{z,m} \rho \sigma_{z,m} )~.
\end{align}
To obtain the solution and the relevant correlation functions, we employ the single-body basis
$
|1\rangle = |g_1 g_2\rangle,  
|2\rangle = |g_1 e_2\rangle,  
|3\rangle = |e_1 g_2\rangle,  
|4\rangle = |e_1 e_2\rangle 
$, where $e_i$ ($g_i$) denotes the $i$-th emitter in the excited (ground) state. 

To describe the physics involving the weakly driven two-photon transition, it is insightful to use the collective coupled eigenstates in the absence of driving. For two coupled emitters detuned by $\delta = \omega_2 - \omega_1$ and in the absence of driving, the coherent terms in the master equation \eqref{eq:master-equation} are diagonalized by the states
\begin{align}
|G\rangle&=|1\rangle, &|E\rangle &= |4\rangle, \\
|S\rangle &= a|3\rangle + b |2\rangle,& |A\rangle &= b|2\rangle - a |3\rangle~,\\
a &= \frac{d}{\sqrt{d^2 + \Omega_{12}^2}}, &b &=\frac{\Omega_{12}}{\sqrt{d^2 + \Omega_{12}^2}}, \\
d &= 2 \delta +\sqrt{4 \delta^2 + \Omega_{12}^2}~.
\end{align}
The corresponding eigenfrequencies
$\omega_E = -\omega_G= (\omega_1+\omega_2)/2$, $\omega_S = - \omega_A = \sqrt{\Omega_{12}^2 + 4 \delta^2}$ provide the level scheme shown in Fig. 1a in the main text and describe well the spectrum encountered at sufficiently weak drivings.
In the case of identical particles $\delta = 0, \gamma_1=\gamma_2$, the coupled basis recovers the bright and dark maximally entangled Dicke states ($|S\rangle , |A\rangle \to  |+\rangle,|-\rangle$) with decay rates $\gamma_{\pm} = \gamma \pm |\gamma_{12}|$, respectively. These states govern the dynamics at weak driving \cite{Ficek2002,Gonzalez-Tudela2011}, whereas in the detuned case $\delta\ne 0$, $|G \rangle$ and $|E\rangle$ are mainly responsible for the response at the two-photon resonance as discussed in the main text.

\subsection{Incoherent fluorescence}
One relevant experimental observable used to identify the two-photon resonance is the incoherent fluorescence as a function of the driving frequency \cite{Hettich2002}, which can be measured outside the spectral range of the exciting laser.
By assessing where the fluorescence level is still low, one can identify the parameters for which coherent effects are expected to be significant. The fluorescence signal is proportional to $I= \sum_i \langle\hat{\vec{E}}^{-}_i \cdot \hat{\vec{E}}^{+}_i\rangle=\sum_{i} |g_i (\vec{r})|^2 \mean{\sigma_i^\dagger \sigma_i}$. In the main text we set $|g_1 (\vec{r}) |=|g_2 (\vec{r})|$, which is a reasonable assumption when the emitters are placed at equivalent positions with respect to the nanostructure.  This can be inferred from Eqs.\,\eqref{eq:field}, \eqref{eq:decay_rate_frequency_shift}, where for similar emitter positions the one-point Green's tensor provides $\boldsymbol{G}(\vec{r}_1,\vec{r}_1, \omega_0) \approx \boldsymbol{G}(\vec{r}_2,\vec{r}_2, \omega_0)$ and at equivalent far-field detection distances $\boldsymbol{G}(\vec{r},\vec{r}_1, \omega_0)  \approx \boldsymbol{G}(\vec{r},\vec{r}_2, \omega_0)$. Nevertheless, we have numerically checked that values differing by up to an order of magnitude between $|g_1 (\vec{r})|^2$ and $|g_2 (\vec{r})|^2$ and between $\gamma_1$ and $\gamma_2$, 
the impact of the two-photon transition remains qualitatively similar to the case shown in Fig. 1 in the main text.
Mainly, this introduces an asymmetry in the fluorescence and a gives a stronger weight in intensity to the emitter with the larger scattering amplitude $|g_i|$.
On the other hand, the states $|G\rangle$ and $|E\rangle$ are still connected by a collective transition at $\Delta = \omega_L - (\omega_1+\omega_2)/2= 0$. This can be seen from the perturbative formula for the population of the excited state deduced from Ref.\,\cite{Varada1992} 

\begin{align}
 & \rho_{44} =  \frac{1}{(\gamma_1+\gamma_2)^2+(2\Delta)^2} 
\times\\ &
 \times 
 \biggl| \frac{2 \Omega_{1} \Omega_{2}  (\omega_1 + \omega_2 - 2 \omega_L) }{(\omega_1 - \omega_L) (\omega_2 - \omega_L)} 
 -  \frac{\Omega_{12} [\Omega_{1}^2 + \Omega_{2}^2]}{(\omega_1 - \omega_L) (\omega_2 - \omega_L)}\biggr|^2,  \nonumber 
 \hspace{1cm} 
\end{align}
which can still provide comparable values for $\gamma_1 \neq \gamma_2$ or asymmetric driving conditions $\Omega_{R,1} \neq \Omega_{R,2}$ . 

\subsection{Concurrence near the two-photon resonance}
The amount of entanglement generated in the coupled system may be quantified for a general quantum state using Wootters's concurrence \cite{Wootters1998}
$
C = \max [\{ 0, \lambda_1 - \lambda_2-\lambda_3-\lambda_4\}],
$
where the $\lambda_n^2$ are the eigenvalues of the operator
\mbox{$
\rho (\sigma_{y,1} \otimes \sigma_{y,2}) \rho^{\star} (\sigma_{y,1} \otimes \sigma_{y,2})
$}
 in descending order. The $^{\star}$ denotes the complex conjugate. 
To gain a better understanding of the entanglement induced by the two-photon resonance, we consider the limit of a weak driving field. In this case, the solution will be close to that of an undriven system, for which exact analytical solutions are available. Here, the density matrix keeps a cross-shaped symmetry \cite{Yu2005} and the concurrence is expressed as 
\begin{align}
C \approx& \max\{0,C_1, C_2\}~ \quad \text{with }\nonumber\\
C_1 =& 2|\rho_{41}| - 2 \sqrt{\rho_{22} \rho_{33}}~,\quad
C_2 =& 2|\rho_{23}| - 2 \sqrt{\rho_{11} \rho_{44}}~.
\end{align}
We have checked for our scheme that $C_1$ applies near the two-photon resonance at weak driving, where the single-photon populations are negligible. This shows that entanglement is due to the driving of the two-photon coherence $\rho_{14}$ connecting the double ground and double excited states. When $|\gamma_{12}|>0$, the asymmetry between the dark and bright states is visible in a dispersive contribution to the concurrence at larger positive detunings, for which then $C_2$ dominates.

\subsection{Electric field quadrature fluctuations}
We identify sub-shot-noise fluctuations in an electric field quadrature component 
$\boldsymbol{\hat{E}}_{\theta}=e^{\imath\theta}\boldsymbol{\hat{E}}^{+}+e^{-\imath\theta}\boldsymbol{\hat{E}}^{-}$
 by negative values of its normally ordered variance \cite{Drummond2010}
\begin{align}
\label{variance_quadratureSup}
&\mean{:\![\Delta\boldsymbol{\hat{E}}_{\theta}(\boldsymbol{r},t)]^{2}\!:} = 
 2 \Re[e^{\imath 2\theta}\mean{[\boldsymbol{\hat{E}}^{+}(\mathbf{r},t)]^2} - e^{\imath 2\theta}
\mean{\boldsymbol{\hat{E}}^{+}(\mathbf{r},t)}^{2}  ] 
\nonumber\\
&
 +2\mean{\boldsymbol{\hat{E}}^{-}(\mathbf{r},t) \boldsymbol{\hat{E}}^{+}(\mathbf{r},t)}
 -2|\mean{\boldsymbol{\hat{E}}^{-}(\mathbf{r},t)}|^2 ~,
\end{align}
where $\boldsymbol{\hat{E}}^{-}=(\boldsymbol{\hat{E}}^{+})^{\dag}$
corresponds to the negative-frequency electric field.
Eq.\,\eqref{eq:field} links the field variance to the emitter correlation functions. We find the electric field fluctuations for one  vector component (index suppressed for clarity)
\begin{widetext}
\begin{align}
\label{squeezing}
\mean{:\![\Delta{\hat{E}}_{\theta}(\mathbf{r},t)]^{2}\!:} 
=~& (\Delta\mathcal{E}_1)^2 +  (\Delta\mathcal{E}_2)^2 +  (\Delta\mathcal{E}_{12})^2~,
\\
 \frac{(\Delta\mathcal{E}_1)^2}{2|g_1|^2} =~ & \rho_{33} + \rho_{44} - |\rho_{13} + \rho_{24}|^2 - \Re[e^{ 2 \imath (\theta+ \phi_{1})} (\rho_{31} + \rho_{42})^2]\label{squeezing1}~,\\
 \frac{(\Delta\mathcal{E}_2)^2}{2|g_2|^2} = ~& \rho_{22} + \rho_{44} - |\rho_{12} + \rho_{34}|^2 - \Re[e^{ 2 \imath (\theta+ \phi_{2})} (\rho_{12} + \rho_{34})^2]\label{squeezing2}~,\\
 \frac{(\Delta\mathcal{E}_{12})^2}{2|g_1 g_2|} =~ &  
 		2\Re[e^{\imath (\phi_{2} - \phi_{1})}\rho_{23} 
		+ e^{-\imath (2 \theta + \phi_{1} + \phi_{2,i})}\rho_{14}] -   4	\Re[e^{ -\imath (\theta+\phi_{1})} (\rho_{13} + \rho_{24})]
 		 \Re[e^{ -\imath (\theta+\phi_{2})} (\rho_{12} + \rho_{34})]~.\label{squeezing3}
\end{align}
\end{widetext}
In the present scheme, two-photon squeezing can be observed best in a balanced detection configuration, where $|g_{1}|^2/|g_{2}|^2\approx 1$. In very unbalanced situations, the signal is dominated by the single emitter terms [Eqs. \eqref{squeezing1} or \eqref{squeezing2}] and reduce the visibility of the cooperative squeezing term [Eq. \eqref{squeezing3}]. Nevertheless, we have confirmed numerically  that the cooperative term can be discerned, as in the case of the incoherent fluorescence, for differences between the emitter's coupling constants up to an order of magnitude. 

At weak driving near the two-photon resonance, we can derive a simple semianalytical formula following closely the approximations provided in the previous subsection for the concurrence. We assume again a cross-shaped density matrix, for which Eq. \eqref{squeezing} reduces to
\begin{align}
\label{eq:squeezingCrossedshape}
&\frac{1}{2}\mean{:\![\Delta{\hat{E}}_{\theta}(\mathbf{r},t)]^{2}\!:} 
	 \approx  |g_{1}|^2(\rho_{33} + \rho_{44})
	 		+ |g_{2}|^2 (\rho_{22} + \rho_{44}) \nonumber\\
			&+2|g_{1} g_{2}| \Re[e^{\imath (\phi_{2} - \phi_{1})}\rho_{23} 
			+ e^{-\imath (2 \theta + \phi_{1}+ \phi_{2})}\rho_{14}]. 
\end{align}
Since the single-photon coherences are small near the two-photon resonance at weak drivings ($|\rho_{23}|\ll |\rho_{14}|$), we can further simplify the squeezing signal. Thus for a balanced situation $|g_{1}|= |g_{2}|=|g|, \phi_{1} = \phi_{2}$ we obtain
\begin{align}
\label{eq:squeezingCrossedshapelimit}
\frac{\mean{:\![\Delta{\hat{E}}_{\theta}(\mathbf{r},t)]^{2}\!:}}{2|g|^2} &\approx\rho_{44} + (1-\rho_{11}) - 2\Re[e^{-2 \imath  \theta'} \rho_{14}] \nonumber\\
&\ge - 2|\rho_{14}|~,
\end{align}
where we have redefined the quadrature angle $\theta' = \theta + (\phi_{1} + \phi_{2})/2$ without loss of generality. Reduced electric field fluctuations are thus traced back to the two-photon coherence $\rho_{41}$ and are suppressed by any depopulation of the ground state and optimized for a quadrature angle such that $- 2\Re[e^{-2 \imath  \theta'} \rho_{14}] = - 2| \rho_{14}|$. 

It is instructive to point out the connection to spin-squeezing, considering the fluctuations of the collective operators 
$S_{x} = \sum_{n} (\sigma_n^\dagger  + \sigma_n)/2$, $S_{y} = \sum_{n} (\sigma_n^\dagger  - \sigma_n)/2\imath$, and $S_{z} = \sum_{n} \sigma_{z,n}$.
For driving near the two-photon resonance, we find that $\langle S_x\rangle \approx \langle S_y\rangle \approx 0$.
Spin squeezing \cite{Saito1997, Hammerer2010} occurs when 
\begin{align}
\xi_x = \frac{2 (\langle S_x^2\rangle-\langle S_x\rangle^2)} {|\langle \vec{S} \rangle|} < 1 ~.
\end{align}
Direct algebraic manipulation shows that this is actually equivalent to the condition
\mbox{$
\mean{:\![\Delta{\hat{S}}]^{2}\!:}= (\langle :S_x^2: \rangle - \langle S_x \rangle^2) <0.
$}
Here,
\begin{align}
4\mean{:\![\Delta{\hat{S}}]^{2}\!:}&=  1 - \rho_{11} + \rho_{44} - 2 \Re(\rho_{14})\nonumber\\
&+ [\rho_{22} 
+ \rho_{33} - 2 \Re(\rho_{23})] ~.
\end{align}
Since the antisymmetric state is not driven, the last paranthesis vanishes recovering an expression that is proportional to Eq.\,\eqref{eq:squeezingCrossedshapelimit} for vanishing quadrature phase and equal values of $g_1 = g_2$.
\bibliography{2015-05-20-TwoPhotonBibliography}

\end{document}